\documentclass[10pt]{article}
\usepackage{amsmath}
\usepackage{amsfonts}

\usepackage{amsmath,amssymb}
\usepackage{graphicx}
\usepackage{color}
\usepackage{url}

\renewcommand{\arraystretch}{1}

\newtheorem{theorem}{Theorem}
\newtheorem{lemma}{Lemma}
\newtheorem{corollary}{Corollary}

\newtheorem{definition}{Definition}
\newtheorem{proposition}{Proposition}
\newtheorem{remark}{Remark}

\newcommand{\RNum}[1]{\uppercase\expandafter{\romannumeral #1\relax}}

\newcommand{\ls}[1]
    {\dimen0=\fontdimen6\the\font\lineskip=#1\dimen0
     \advance\lineskip.5\fontdimen5\the\font
     \advance\lineskip-\dimen0
     \lineskiplimit=0.9\lineskip
     \baselineskip=\lineskip
     \advance\baselineskip\dimen0
     \normallineskip\lineskip\normallineskiplimit\lineskiplimit
     \normalbaselineskip\baselineskip
     \ignorespaces}

\begin{document}

\bibliographystyle{abbrv}

\title{A New Class of Permutation Trinomials Constructed from Niho Exponents}

\author{Tao Bai 
	and Yongbo Xia\thanks{Corresponding author. 
			T. Bai and Y. Xia are with the Department of Mathematics and Statistics, South-Central University
			for Nationalities, Wuhan 430074, China. Y. Xia is also with the Hubei Key Laboratory of Intelligent Wireless Communications,
  South-Central University for Nationalities, Wuhan 430074, China (e-mail:
xia@mail.scuec.edu.cn).}
}
\date{}
\maketitle

\thispagestyle{plain} \setcounter{page}{1}

\begin{abstract}
Permutation polynomials over finite fields are an interesting subject due to their important applications in the areas of mathematics and engineering. In this paper we investigate the trinomial $f(x)=x^{(p-1)q+1}+x^{pq}-x^{q+(p-1)}$ over the finite field $\mathbb{F}_{q^2}$, where $p$ is an odd prime and $q=p^k$ with $k$ being a positive integer. It is shown that when $p=3$ or $5$,  $f(x)$ is a permutation trinomial of $\mathbb{F}_{q^2}$ if and only if $k$ is even. This property is also true for  more general class of polynomials $g(x)=x^{(q+1)l+(p-1)q+1}+x^{(q+1)l+pq}-x^{(q+1)l+q+(p-1)}$, where $l$ is a nonnegative integer and $\gcd(2l+p,q-1)=1$.  Moreover, we also show that for $p=5$ the permutation trinomials $f(x)$ proposed here are new in the sense that they are not multiplicative equivalent to previously known ones of similar form.

\medskip

{\bf Index Terms } Finite fields, Permutation polynomials, Trinomials, Niho exponents, Multiplicative inequivalent.

\smallskip

{\bf AMS } 94B15, 11T71
\end{abstract}

\ls{1.5}
\section{Introduction}\label{sec1}

Let $\mathbb{F}_{q}$ denote the finite field with $q$ elements and $\mathbb{F}_{q}^*=\mathbb{F}_{q}\setminus \{0\}$, where $q$ is a prime power.  A polynomial $f(x)\in\mathbb{F}_q[x]$ is called a permutation polynomial of $\mathbb{F}_q$ if the associated mapping $f:\,c\longmapsto f(c)$ permutes $\mathbb{F}_{q}$. Permutation polynomials were firstly studied by Hermite for the finite prime fields and by Dickson for arbitrary finite fields \cite{LN}. They have wide applications in coding theory \cite{Laigle-Chapuy2007}, cryptography \cite{Dobbertin1999,Qu2013} and combinatorial designs \cite{Ding and Yua2006}. For a finite field $\mathbb{F}_q$, there are in total $q!$ permutation polynomials of $\mathbb{F}_q$, and all of them can be obtained from the Lagrange interpolation. %However, for applicative purposes, it is usually required that the permutation polynomials have a simple or nice appearance or possess additional extraordinary properties. For this reason, people are mainly interested in the methods for constructing permutation polynomials with few terms. 
Permutations with a few terms are of particular interest because of the simple algebraic expressions. Especially, permutation binomials and trinomials attracted particular attention \cite{Hou2015survey, X.Hou2016arix,Likangquan,Tu-ZengFFA,Nian-Tor-FFA,Tu-ZengFFA2015}.  Recent achievements on the study of permutation polynomials were surveyed in \cite{Hou2015survey,Handbook2013}.

Let $p$ be a prime and $k$  a positive integer. A  Niho exponent over the finite field $\mathbb{F}_{p^{2k}}$ is a positive integer $d$ satisfying $d\equiv p^j\,\,({\rm mod}\,\,p^k-1)$ for some nonnegative integer $j<k$. In the case of $j=0$, it is called a normalized Niho exponent. Researches in the past decades  demonstrate that Niho exponents are good resources that lead to desirable objects in sequence design \cite{Dobbertin06,xiaieee2016}, coding theory \cite{chenyuan2016,zengcyclic2010} and cryptography \cite{Niho-BENT-Dobbertin06}. 
Recently, a lot of permutation trinomials of the form 
\begin{equation}\label{niho0-per}F(x)=x+\lambda_1 x^{s(p^k-1)+1}+\lambda_2 x^{t(p^k-1)+1}\end{equation}
have been proposed, where $s$ and $t$ are two  integers,  and the coefficients $\lambda_1$ and $\lambda_2$ are restricted to $\{-1,\,1\}$. For $p=2$, Li and Helleseth gave a rather detailed list of known pairs $(s,t)$ and some new pairs such that $F(x)$  is a permutation polynomial of $\mathbb{F}_{2^{2k}}$ \cite{Linian2016cc,Linian2016jun}. In \cite{Likangquan2016arxiv,Gupta2016,zhazhengbang2017,wudanyao2017} some permutation trinomials of $\mathbb{F}_{2^{2k}}$ of similar form were also presented.
For $p=3$, Li \textit{et al.} in \cite{Likangquan2016arxiv} investigated several permutation trinomials of $\mathbb{F}_{3^{2k}}$ of the form $(\ref{niho0-per})$ and proposed three conjectures, which were later confirmed in \cite{Linian2016conjecture} and \cite{Bartoli2017}. Very recently, for $p=5$, Wu and Li in \cite{Wu-Li2017} derived a series of sufficient conditions on $s$, $t$, $\lambda_1$ and $\lambda_2$ for $F(x)$ 
to permute $\mathbb{F}_{5^{2k}}$. 

%There are also some permutation polynomials over more %general finite fields that are constructed from Niho %exponents. 
There are also some permutation polynomials constructed from Niho exponents over $\mathbb{F}_{q^2}$ with  $q$ being a power of an arbitrary prime. 
Hou in \cite{X.Hou-2014arxiv} characterized the necessary and sufficient conditions on the coefficients for the polynomial $ax+bx^q+x^{2q-1}\in \mathbb{F}_{q^2}[x]$ to be a permutation of $\mathbb{F}_{q^2}$. 
%In \cite{X.Hou-2014method}, for two given Niho %exponents $q$ and $2q-1$ over $\mathbb{F}_{q^2}$,  a %class of permutation polynomials of $\mathbb{F}_{q^2}$ %of the form $ax+bx^q+x^{2q-1}\in \mathbb{F}_{q}[x]$ %was fully studied, and the explicit conditions on $a$ %and $b$ that are necessary and sufficient for %$ax+bx^q+x^{2q-1}$ to be a permutation of %$\mathbb{F}_{q^2}$ are found. 
In \cite{Ding-siam}, for $q\not\equiv 3\,\,(\rm{mod}\,\,3)$, the necessary and sufficient conditions for $x+x^{t(q-1)+1}+x^{-t(q-1)+1}$ to be a permutation polynomial of $\mathbb{F}_{q^2}$ were determined, where $t$ is a positive integer. Let  $Tr_{q^2/q}(\cdot)$ denote the trace function from $\mathbb{F}_{q^2}$ to $\mathbb{F}_{q}$ \cite{LN}. Some permutation trinomials of $\mathbb{F}_{q^2}$ of the form  $x+\gamma Tr_{q^2/q}(x^d)$  were obtained in \cite{Kyureghyan-Zieve2016}, where $\gamma\in\mathbb{F}_{q}^*$ and $d$ is a Niho exponent over $\mathbb{F}_{q^2}$. 
%Note that the above are just some sparse permutation %polynomials constructed from Niho exponents, and it is %still of interest to construct new permutation %polynomials from Niho exponnets.  

  In this paper, we investigate the permutation property of the following trinomial 
\begin{equation}\label{niho4-per}f(x)=x^{(p-1)q+1}+x^{pq}-x^{q+(p-1)},\end{equation}
 where $p$ is an odd prime and $q=p^k$ for a positive integer $k$. It is easily verified that $(p-1)q+1$, $pq$ and $q+(p-1)$ are Niho exponents over $\mathbb{F}_{p^{2k}}$.   We show that for $p=3$ or $5$, $f(x)$ in (\ref{niho4-per}) is a permutation polynomial of $\mathbb{F}_{p^{2k}}$ if and only if $k$ is even. However, for the case $p>5$, such a result may not hold. Furthermore, we prove that the above property is also true for more general polynomials
\begin{equation}\label{general-niho-per}g(x)=x^{(q+1)l+(p-1)q+1}+x^{(q+1)l+pq}-x^{(q+1)l+q+(p-1)},\end{equation} where $l$ is a nonnegative integer and $\gcd(2l+p,q-1)=1$.
In addition, when $p=5$, the permutation polynomials $f(x)$ presented in  (\ref{niho4-per}) are shown to be new in the sense that they are not multiplicative equivalent to the permutation polynomials of the form (\ref{niho0-per}) in \cite{Likangquan2016arxiv,Linian2016conjecture,Wu-Li2017,mage2017,X.Hou-2014method,X.Hou-2014arxiv,Ding-siam,Kyureghyan-Zieve2016}.

The remainder of this paper is organized as follows. Section
\ref{pre} gives some preliminaries and notation, including some
useful lemmas. In Section \ref{sec3}, we give the proofs of our main results. Section \ref{sec4} is devoted to demonstrating that the permutation trinomials $f(x)$ given in  (\ref{niho4-per}) are new when $p=5$. Section \ref{sec5} concludes the study.

\section{Preliminaries}\label{pre}

Let $p$ be a prime, $k$  a positive integer and $q=p^k$. The trace function and the norm function from $\mathbb{F}_{q^{2}}$ to $\mathbb{F}_{q}$ will be denoted by $Tr(x)$ and $N(x)$, respectively \cite{LN}. Namely, $$Tr(x)=x+x^q\,\,\mbox{ and }\,\,N(x)=x\cdot x^q,\,\, x\in\mathbb{F}_{q^{2}}.$$
The unit circle $U$ of $\mathbb{F}_{q^{2}}$ is defined by 
\begin{equation}\label{defU}
U=\{x\,\mid\,x^{q+1}=1,\,\,x\in \mathbb{F}_{q^{2}}\}.
\end{equation}

In \cite{Linian2016cc,Linian2016jun,Likangquan2016arxiv,Linian2016conjecture,Wu-Li2017}, in order to prove the permutation property of the trinomials constructed from Niho exponents, the authors mainly used the following lemma, which was proved by Park and Lee in 2001 and reproved by Zieve in 2009.

\begin{lemma}\label{zieve lemma}(\cite{Park2001,Zieve2009}) Let $p$ be a prime and $n$  a positive integer. Assume that $d$ is a positive integer such
that $d\mid (p^n-1)$, $h(x)\in \mathbb{F}_{p^n}[x]$ and $r>0$ is a integer. Then, $x^rh(x^{\frac{p^n-1}{d}})$ is a permutation of 
$\mathbb{F}_{p^n}$ if and only if 

\noindent (i) $\gcd(r,\frac{p^n-1}{d})=1$ and

\noindent (ii) $x^rh(x)^{\frac{p^n-1}{d}}$ permutes $\mu_d$,
where $\mu_d$ is the set of $d$-th root of unity in $\mathbb{F}_{p^n}^*$.
\end{lemma}

The polynomials constructed from Niho exponents over $\mathbb{F}_{q^{2}}$ can always be rewritten as the form $x^rh(x^{\frac{q^2-1}{d}})$ with $d=q+1$. To determine the permutation property of the polynomials $x^rh(x^{\frac{q^2-1}{d}})$ constructed from Niho exponents by Lemma \ref{zieve lemma}, the main task is to  decide if  $x^rh(x)^{\frac{q^2-1}{d}}$ permutes the unit circle $U$ of $\mathbb{F}_{q^2}$.  However, sometimes the corresponding polynomial $x^rh(x)^{\frac{q^2-1}{d}}$ leads to fractional polynomial with high degree \cite{Tu-Zeng-Hu-Li,Linian2016cc,Linian2016jun,Likangquan2016arxiv,Linian2016conjecture,Wu-Li2017}. It is still a difficult problem in general to verify that   $x^rh(x)^{\frac{q^2-1}{d}}$ permutes $U$. 

Another general approach to investigating the permutation property of the polynomials constructed from Niho exponents over $\mathbb{F}_{q^{2}}$ is to concentrate on the subset $\mathbb{F}_{q^2}\setminus \mathbb{F}_q$. More specifically,  assume that $G(x)$ is a polynomial constructed from Niho exponents over $\mathbb{F}_{q^2}$ with coefficients in $\mathbb{F}_q$. If we can show that $G(x)$ is a permutation of $\mathbb{F}_q$ and a permutation of $\mathbb{F}_{q^2}\setminus \mathbb{F}_q$ respectively, then $G(x)$ is 
a permutation polynomial of $\mathbb{F}_{q^2}$. To this end, it is usually required that $G(x)$ has the property $G(\mathbb{F}_{q^2}\setminus \mathbb{F}_q)\in \mathbb{F}_{q^2}\setminus \mathbb{F}_q$, and the key step in the proof is to prove that $G(x)$ is a permutation of $\mathbb{F}_{q^2}\setminus \mathbb{F}_q$. This idea originated from \cite{X.Hou-2014method} and later was used in \cite{Likangquan2016arxiv}. In this paper we will use this idea to prove our main result. 

The following lemma is needed in the sequel. Its proof is trivial and is omitted here.

\begin{lemma}\label{N-T determined} Let $q$ be a prime  power. Denote by $Tr(x)$ and $N(x)$ the trace function and the norm function from $\mathbb{F}_{q^2}$ to $\mathbb{F}_{q}$, respectively. Then, for any $c\in \mathbb{F}_{q^2}$, $\{c,c^q\}$ is uniquely determined by the pair $(Tr(c),N(c))$.
\end{lemma}

The following lemma is obtained by direct computations. 

\begin{lemma}\label{N-T express}Let $q$ be a prime power, and $Tr(x)$ and $N(x)$ be the trace function and the norm function from $\mathbb{F}_{q^2}$ to $\mathbb{F}_{q}$, respectively.

 \noindent (i) If $q=3^k$, then for any $x\in \mathbb{F}_{q^2}$,
 $$Tr(x^2)=Tr^2(x)+N(x)\,\,\mbox{and}
   \,\,Tr(x^4)=Tr^4(x)-N(x)Tr^2(x)-N^2(x);$$
 
 \noindent (ii) If $q=5^k$, then for any $x\in \mathbb{F}_{q^2}$,
  $$Tr(x^2)=Tr^2(x)-2N(x),\,\,Tr(x^3)=Tr^3(x)+2N(x)Tr(x),$$
  $$Tr(x^4)=Tr^4(x)+N(x)Tr^2(x)+2N^2(x),$$
  $$Tr(x^6)=Tr^6(x)-N(x)Tr^4(x)-N^2(x)Tr^2(x)-2N^3(x),$$
  and 
  $$Tr(x^8)=Tr^8(x)+2N(x)Tr^6(x)-N^3(x)Tr^2(x)+2N^4(x).$$
 
\end{lemma}
{\em Proof:} We only give the proof of \noindent (ii). In Lemma 1 of \cite{Wu-Li2017}, the expressions for $Tr(x^2)$, $Tr(x^3)$, $Tr(x^4)$ and $Tr(x^8)$ were given while that for $Tr(x^6)$ was not. Now we compute $Tr(x^6)$ to illustrate how to obtain the above results. Note that
\begin{equation*}
\begin{array}{lll}Tr^6(x)&=&(x+x^q)^6\\
&=&(x+x^q)^5(x+x^q)\\
&=&(x^5+x^{5q})(x+x^q)\\
&=&x^6+x^{5+q}+x^{5q+1}+x^{6q}\\
&=&Tr(x^6)+x^{1+q}\left(x^{4}+x^{4q}\right)\\
&=&Tr(x^6)+N(x)Tr(x^4),\\
\end{array}
\end{equation*}
which implies \begin{equation}\label{Trx6}
Tr(x^6)=Tr^6(x)-N(x)Tr(x^4).
\end{equation} 
Substituting $Tr(x^4)$ into (\ref{Trx6}), we get the desired result.
\hfill$\square$

\begin{lemma}\label{root of quad} Let $U$ be defined as (\ref{defU}). We have the following results:

 \noindent (i) if $q=3^k$, then $y^2+1=0$ has no root in $U$ if $k$ is even and has two roots in $U$ otherwise;

 \noindent (ii) if $q=5^k$, then $y^2-y+1=0$ has no root in $U$ if $k$ is even and has two roots in $U$ otherwise.
\end{lemma}
{\em Proof:} (i) Let $\alpha$ be a primitive root of $\mathbb{F}_{q^2}$ with $q=3^k$. Then, the roots of $y
^2=-1$ are $\pm \alpha^{\frac{q^2-1}{4}}$, which belong to $U$ if and only if $q+1$ is 
divisible by $4$. Since $q+1$ is 
divisible by $4$ if and only if $k$ is odd, it follows the desired result. 

(ii) Note that $y^2-y+1$ is an irreducible polynomial over $\mathbb{F}_5$ because it has no solution in $\mathbb{F}_5$. Since the degree of $y^2-y+1$ is $2$, it follows that its two roots belong to $\mathbb{F}_{5^2}$. We can rewrite  $y^2-y+1$ as $(y+2)^2-3$. Then, the two roots of $y^2-y+1$ in $\mathbb{F}_{5^2}$ are $-2\pm \sqrt{3}$, where $\pm\sqrt{3}$ denote the two roots of $x^2=3$ in $\mathbb{F}_{5^2}$. 

When $k$ is even, $\left(\sqrt{3}\right)^q=\sqrt{3}$ since $\sqrt{3}\in \mathbb{F}_{5^2}$ and when $k$ is odd,  $\left(\sqrt{3}\right)^q=\left(\sqrt{3}\right)^5=-\sqrt{3}$. Thus, when $k$ is even, we have 
\begin{equation*}
\begin{array}{lll}\left(-2\pm \sqrt{3}\right)^{q+1}&=&\left(-2\pm \sqrt{3}\right)^q\left(-2\pm \sqrt{3}\right)\\
&=&\left(-2\pm \sqrt{3}\right)^2\\
&=&2\pm\sqrt{3},\\
\end{array}
\end{equation*}
which is not equal to $1$. When $k$ is odd, we have 
\begin{equation*}
\begin{array}{lll}\left(-2\pm \sqrt{3}\right)^{q+1}&=&\left(-2\mp \sqrt{3}\right)\left(-2\pm \sqrt{3}\right)\\
&=&4-3\\
&=&1.\\
\end{array}\end{equation*}Therefore, when $k$ is odd,  $-2\pm \sqrt{3}\in U$. From the above computations, it follows the desired result. 
\hfill$\square$

The following lemma is a special case of Exercise 7.4 in \cite{LN}. For the reader's convenience, we include a proof here. 
\begin{lemma}\label{permutation 5} Let $\mathbb{F}_q$ be a finite field of characteristic $p$. Then, $x^p-ux\in \mathbb{F}_q[x]$ is a permutation polynomial of $\mathbb{F}_q$ if and only if $u$ is not a $(p-1)$th power of an element of $\mathbb{F}_q^*$.
\end{lemma}
{\em Proof:} Note that  $x^p-ux$ is a $p$-polynomial over $\mathbb{F}_q$, and it is a permutation polynomial of $\mathbb{F}_q$ if and only if it only has the root $0$ in $\mathbb{F}_q$. Thus, $x^p-ux$ is a permutation polynomial of $\mathbb{F}_q$ if and only if 
$x^{p-1}-u$ has no root in $\mathbb{F}_q$. The latter exactly means that $u$ is not a $(p-1)$th power in $\mathbb{F}_q$. 
 \hfill$\square$

\section{A new class of permutation trinomials from Niho exponents}\label{sec3}

In this section, we present our main results about the permutation property of $f(x)$ defined in (\ref{niho4-per}) and that of $g(x)$ defined in (\ref{general-niho-per}). The first main result is given in the following theorem. 

\begin{theorem}\label{main result}
Let $q=p^k$ and $f(x)$ be the trinomial defined in (\ref{niho4-per}). 
Then for $p=3$ or $5$, $f(x)$ is a permutation polynomial of $\mathbb{F}_{q^{2}}$ if and only if $k$ is even.
\end{theorem}

Before we prove this theorem in detail, we mention some characterizations of $f(x)$ and $g(x)$. The three exponents appearing in $f(x)$ are Niho exponents since $(p-1)q+1=(p-1)(q-1)+p$, $pq=p(q-1)+p$, and $q+(p-1)=(q-1)+p$. However, the exponents in $g(x)$ may not be Niho exponents. As we will see in the sequel, the permutation property of $f(x)$ and that of $g(x)$ depend on a same condition after utilizing Lemma \ref{zieve lemma}. If we obtain the condition for $f(x)$ to permute $\mathbb{F}_{q^2}$, then it is also true for $g(x)$. In addition,  a permutation polynomial $f(x)$ of the form (\ref{niho4-per}) is closely related to  some permutation polynomial of the form  (\ref{niho0-per}). In the next section, we will study the relationship between $f(x)$ in Theorem \ref{main result} and the previously known permutation trinomials of the form (\ref{niho0-per}). By comparison, we will show that the permutation polynomial $f(x)$ proposed here is new when $p=5$.

%A permutation polynomial with form (\ref{niho4-per}) sometimes can be %multiplicative equivalent to some permutation of the form   %(\ref{niho3-per}). In next section, we will show the permutation %trinomial $f(x)$ in Theorem \ref{main result} is not multiplicative %equivalent to any one presented in \cite{Wu-Li2017}. In this sense we %indeed obtain a new class of permutation trinomials over %$\mathbb{F}_{5^{2k}}$.

 In order to prove Theorem \ref{main result}, the following preparatory lemma is needed. 

\begin{lemma}\label{pre lemma} Let $q=p^k$ and $f(x)$ be defined in (\ref{niho4-per}). When $p=3$ or $5$, for any $x\in \mathbb{F}_{q^2}\setminus \mathbb{F}_q$, we have $f(x)\in\mathbb{F}_{q^2}\setminus \mathbb{F}_q $ if and only  if $k$ is even.
\end{lemma}
{\em Proof:} Assume that  $x\in \mathbb{F}_{q^2}\setminus \mathbb{F}_q$.  Note that $f(x)\in \mathbb{F}_q $ if and only if 
$$x^{(p-1)q+1}+x^{pq}-x^{q+(p-1)}=x^{q+(p-1)}+x^p-x^{(p-1)q+1}$$
which is equivalent to 
\begin{equation*}\label{prefor main 1}
2x^{q+(p-1)}+x^p-x^{pq}-2x^{(p-1)q+1}=0.
\end{equation*}
Dividing the above equation by $x^p$, we get 
\begin{equation}\label{prefor main 2}
2x^{q-1}+1-x^{p(q-1)}-2x^{(p-1)(q-1)}=0.
\end{equation}
Setting $z=x^{q-1}$, (\ref{prefor main 2}) can be rewritten as 
\begin{equation*}\label{prefor main 3a}2z+1-z^p-2z^{p-1}=0,\end{equation*}
which equals 
\begin{equation}\label{prefor main 3b}
(z-1)(z^2+1)=0
\end{equation}
 if $p=3$, and 
\begin{equation}\label{prefor main 3}
(z-1)(z^2-z+1)^2=0
\end{equation}
if $p=5$.

Note that if $x\in \mathbb{F}_{q^2}\setminus \mathbb{F}_q$, then $z=x^{q-1}\in U\setminus \{ 1\}$, where $U$ is defined as (\ref{defU}).  For $p=3$, from (\ref{prefor main 3b}), we can conclude that for any $x\in \mathbb{F}_{q^2}\setminus \mathbb{F}_q$, we have $f(x)\in \mathbb{F}_q $ if and only if $z=x^{q-1}$ satisfies $z^2+1=0$. By Lemma \ref{root of quad}, if $k$ is even, then $z^2+1=0$
has no root in $U$. Thus, in this case, for any $x\in \mathbb{F}_{q^2}\setminus \mathbb{F}_q$, we have $f(x)\notin\mathbb{F}_q $, \textit{i.e.}, $f(x)\in\mathbb{F}_{q^2}\setminus \mathbb{F}_q$. If $k$ is odd, then $z^2+1=0$ has two roots in $U\setminus \{ 1\}$, which shows that there are  $2(q-1)$ elements 
$x\in\mathbb{F}_{q^2}\setminus \mathbb{F}_q$ such that $f(x)\in \mathbb{F}_q$.  Similarly, for $p=5$, we can obtain the same conclusion by (\ref{prefor main 3}) and Lemma \ref{root of quad}.

From the above discussions, it follows the desired result. 
\hfill$\square$

{\em \textbf{Proof of Theorem \ref{main result}}.} Note that in this proof $q=p^k$ and $p$ is restricted to $\{3,5\}$. To prove that $f(x)$ is a permutation of $\mathbb{F}_{q^2}$, it suffices to show that for any $c\in \mathbb{F}_{q^2}$, $f(x)=c$ has exactly one root in $\mathbb{F}_{q^2}$.

Note that when $x\in \mathbb{F}_q$, then $f(x)=x^p$, which is a permutation of $\mathbb{F}_q$ since $\gcd(p,q-1)=1$. On the other hand, if $k$ is odd, from the proof of Lemma \ref{pre lemma}, there are  $2(q-1)$ elements $x\in\mathbb{F}_{q^2}\setminus \mathbb{F}_q$ such that $f(x)\in \mathbb{F}_q$. Therefore, when $k$ is odd, for some $c\in \mathbb{F}_q$ there must exist at least two distinct elements  $ x_1,\,x_2\in\mathbb{F}_{q^2}$ such that $f(x_1)=f(x_2)=c$. Thus, $f(x)$ is not a permutation polynomial of $\mathbb{F}_{q^2}$ when $k$ is odd.

Next we prove that when $k$ is even, then $f(x)$ is a permutation polynomial of $\mathbb{F}_{q^2}$, \textit{i.e.}, $f(x)=c$ has exactly one root in $\mathbb{F}_{q^2}$. We consider the following two cases. 

\emph{Case 1:} $c\in \mathbb{F}_q$. Then, by Lemma \ref{pre lemma}, we can derive from $f(x)=c$ that $x$ must belong to $\mathbb{F}_q$. Thus, in this case $f(x)=c$ is equivalent to that $x^p=c$. Obviously, the latter has only one root in $\mathbb{F}_{q^2}$.

\emph{Case 2:} $c\in \mathbb{F}_{q^2}\setminus\mathbb{F}_q$. Then, by Lemma \ref{pre lemma}, the roots of $f(x)=c$ belong to $\mathbb{F}_{q^2}\setminus\mathbb{F}_q$. Next we will show that $f(x)=c$ has exactly one root in $\mathbb{F}_{q^2}\setminus\mathbb{F}_q$ under the given conditions.  Let $Tr(x)$ and $N(x)$ be defined as (\ref{N-T express}). We compute $Tr(f(x))$ and $N(f(x))$ as follows:
\begin{equation}\label{tf}\begin{array}{lll}
Tr(f(x))&=&\left(x^{(p-1)q+1}+x^{pq}-x^{q+(p-1)}\right)+\left(x^{(p-1)q+1}+x^{pq}-x^{q+(p-1)}\right)^q\\
&=&x^{pq}+x^p\\
&=&Tr^p(x),
\end{array}
\end{equation}
and 
\begin{equation}\label{nf}\begin{array}{lll}
N(f(x))&=&\left(x^{(p-1)q+1}+x^{pq}-x^{q+(p-1)}\right)\left(x^{q+(p-1)}+x^{p}-x^{(p-1)q+1}\right)\\

&=&3x^{pq+p}+\left(x^{(p-1)q+(p+1)}+x^{(p+1)q+(p-1)}\right)\\
&&-\left(x^{(2p-2)q+2}+x^{2q+(2p-2)}\right)-\left(x^{(2p-1)q+1}+x^{q+(2p-1)}\right)\\
&=&3N^p(x)+N^{p-1}(x)Tr(x^2)-N^2(x)Tr(x^{2p-4})-N(x)Tr(x^{2p-2}).
\end{array}
\end{equation}
When $p=3$ or $5$, by Lemma \ref{N-T express}, $N(f(x))$ in (\ref{nf}) can be expressed in terms of $Tr(x)$ and $N(x)$ as follows:
\begin{equation}\label{nfs}\arraycolsep=1.2pt\def\arraystretch{1.7}
N(f(x))=\left \{\begin{array}{lll}
-N(x)Tr^{4}(x)+N^2(x)Tr^2(x)+N^3(x),\,\,\mbox{for}\,\,p=3,\\
N^5(x)+3N^4(x)Tr^2(x)+N^3(x)Tr^4(x)\\
~~~~~~~~~+2N^2(x)Tr^6(x)-N(x)Tr^8(x),\,\,\mbox{for}\,\,p=5.
\end{array}\right.\end{equation}
For any $c\in \mathbb{F}_{q^2}\setminus \mathbb{F}_{q}$, from $f(x)=c$, we have 

\begin{equation}\label{TNc}Tr(f(x))=Tr(c)\,\,\mbox{and}\,\,N(f(x))=N(c).\end{equation} In what follows, we will prove that $Tr(x)$ and $N(x)$ are uniquely determined by $c$ under the aforementioned conditions. We only give the proof of this conclusion in the case $p=5$, and for $p=3$ it can be proved in the same way.  Thus, in the sequel we always assume that $q=5^k$.

By (\ref{tf}), (\ref{nfs}) and (\ref{TNc}), we have 
\begin{equation}\label{system equation} \arraycolsep=1.2pt\def\arraystretch{1.7}
\left \{\begin{array}{lll}
Tr^5(x)=Tr(c),\\
N^5(x)+3N^4(x)Tr^2(x)+N^3(x)Tr^4(x)+2N^2(x)Tr^6(x)-N(x)Tr^8(x)=N(c).
\end{array}\right.\end{equation}
Note that $\gcd(5,q-1)=1$. Thus, from $Tr^5(x)=Tr(c)$, one knows that $Tr(x)$ is uniquely determined by $c$.
Therefore, it suffices to show that $N(x)$ is also uniquely determined by $c$. We consider the following two subcases. 

\emph{Subcase 2.1:} $Tr(c)=0$. Then, it follows that $Tr(x)=0$. From the second equation in (\ref{system equation}), we have 
$N^5(x)=N(c)$ and thus $N(c)$ is also uniquely determined by $c$.

\emph{Subcase 2.2:} $Tr(c)\neq 0$. Then, $Tr(x)\neq 0$. For convenience, put \begin{equation}\label{define r s} r=\frac{N(x)}{Tr^2(x)}\,\,\mbox{and}\,\,s={N(c)\over Tr^2(c)}.\end{equation} Then, divided by $Tr^{10}(x)$, the second equation in (\ref{system equation}) can be rewritten as 
\begin{equation*}\label{final 1}
r^5+3r^4+r^3+2r^2-r=s
\end{equation*}
or equivalently 
\begin{equation}\label{final 1}
(r-2)^4(r+1)=s+1.
\end{equation}
We claim  that $s$ is not equal to $-1$. Otherwise, we have $N(c)+Tr^2(c)=0,$
which implies $\left(c^q-c\right)^2=0$ leading to $c\in \mathbb{F}_{q}$, a contradiction to the assumption $c\in \mathbb{F}_{q^2}\setminus \mathbb{F}_q$. Let \begin{equation}\label{deft}t=r-2,\end{equation} then (\ref{final 1}) becomes
\begin{equation}\label{final 2}t^5+3t^4=s+1.\end{equation} Note that $t$ is not equal to  $0$ since $s$ is not equal to $-1$. Therefore, (\ref{final 2}) can be further transformed into 
\begin{equation}\label{final 3}\left(\frac{1}{t}\right)^5-\frac{3}{s+1}\cdot\frac{1}{t}=\frac{1}{s+1}.\end{equation} 
If we can show that $\frac{3}{s+1}$ is not a fourth power in $\mathbb{F}_q$, then by (\ref{final 3})
and Lemma \ref{permutation 5} we can conclude that $\frac{1}{t}$ is uniquely determined by $s$. Then, from  (\ref{define r s}) and (\ref{deft}), it follows that $N(x)$ is uniquely determined by $c$. 

Thus, it suffices to show that $\frac{3}{s+1}$ is not a fourth power in $\mathbb{F}_q$. To this end, we express $\frac{3}{s+1}$ in terms of $c$ as follows:
\begin{equation}\label{fourthpower}
\frac{3}{s+1}=\dfrac{3}{N(c)/Tr^2(c)+1}=3\left(1-\frac{1}{c^{1-q}+c^{q-1}-2}\right).
\end{equation}
Since $c\in \mathbb{F}_{q^2}\setminus \mathbb{F}_{q}$ and $Tr(c)\neq 0$, it follows that $c^{q-1}\in U\setminus \{\pm1\}$. Let $u=c^{q-1}$. Suppose, on the contrary, that $\frac{3}{s+1}$ is a fourth power in $\mathbb{F}_q$. Then, by (\ref{fourthpower})  we have 
\begin{equation}\label{fourthpower1}
3\left(1-\frac{1}{u+u^{-1}-2}\right)=\omega^4
\end{equation}
 for some $\omega\in \mathbb{F}_q$. Note that $3-\omega^4\neq 0$. Thus, we can rewrite (\ref{fourthpower1}) as 
 \begin{equation*}\label{fourthpower2}
u^2+\dfrac{1+2\omega^4}{3-\omega^4}\cdot u+1=0
 \end{equation*}
which implies 
\begin{equation}\label{fourthpower3}
u=\dfrac{3\omega^4+4\pm\sqrt{3}\omega^2}{2(3-\omega^2)},
 \end{equation}
where $\pm\sqrt{3}$ denote the two root of $x^2=3$ in $\mathbb{F}_{5^2}$. Note that when $k$ is even, the two roots of $x^2=3$, which are $\pm\sqrt{3}$, belong to $\mathbb{F}_q$. From (\ref{fourthpower3}), it follows that $u\in \mathbb{F}_q$, which contradicts to  $u\in U\setminus \{\pm1\}$ since $\mathbb{F}_q\cap U=\{\pm1\}$. Therefore, $\frac{1}{s+1}$ is not a fourth power in $\mathbb{F}_q$ and the desired result follows.

The discussions in Subcases 2.1 and 2.2 show that for $p=5$, when $k$ is even and $c\in \mathbb{F}_{q^2}\setminus\mathbb{F}_q$,  we can derive from $f(x)=c$ that $Tr(x)$ and $N(x)$ are uniquely determined by $c$. For $p=3$, this conclusion can be similarly proved.

Furthermore, by Lemma \ref{N-T determined}, when $p\in \{3,5\}$, for any $c\in \mathbb{F}_{q^2}\setminus\mathbb{F}_q$, one can conclude from $f(x)=c$ that the set $\{x,x^q\}$ is uniquely determined by $c$. Note that if one of $\{x,x^q\}$ satisfies $f(x)=c$, the other does not. For instance, if $f(x)=c$, then $f(x^q)=\left(f(x)\right)^q=c^q\neq c$ since $c\in \mathbb{F}_{q^2}\setminus\mathbb{F}_q$. Thus, for $p=3$ or $5$, when $k$ is even, for any $c\in \mathbb{F}_{q^2}\setminus\mathbb{F}_q$, $f(x)=c$ has only one root in $\mathbb{F}_{q^2}\setminus\mathbb{F}_q$.

From Cases 1 and 2, it follows the desired result. \hfill$\square$

\begin{corollary}\label{coro1}
Let $U$ be defined as (\ref{defU}) with $q=p^k$ and $p\in \{3,5\}$. Then, the following fractional polynomial
\begin{equation}\label{fractional}
\frac{x+1-x^{p-1}}{x^p+x^{p-1}-x}
\end{equation}
permutes $U$ if and only if $k$ is even.

\end{corollary}
{\em Proof:} Note that $f(x)$ in (\ref{niho4-per}) can be written as $x^p\left(x^{(p-1)(q-1)}+x^{p(q-1)}-x^{q-1}\right)$.
Then, by Lemma \ref{zieve lemma} and Theorem \ref{main result}, it follows that $x^p\left(x^{p-1}+x^p-x\right)^{q-1}$ permutes $U$ if and only if $k$ is even, where $p\in \{3,5\}$. Note that if $x^p\left(x^{p-1}+x^p-x\right)^{q-1}$ permutes $U$, it is not equal to zero when $x\in U$. Thus, for each $p\in \{3,5\}$, $x^p\left(x^{p-1}+x^p-x\right)^{q-1}$ can be written as 
$$\frac{x^p\left(x^{p-1}+x^p-x\right)^q}{x^{p-1}+x^p-x},$$
which is exactly (\ref{fractional}) since $x^q=x^{-1}$ when $x\in U$. 
 \hfill$\square$
 
 Based on Lemma \ref{zieve lemma} and Corollary \ref{coro1}, we obtain our second main result which gives the permutation property of $g(x)$ defined in (\ref{general-niho-per}).
\begin{theorem}\label{permutation property of g(x)}
Let $q=p^k$ with $p\in \{3,5\}$, and $l$  a nonnegative integer satisfying $\gcd(2l+p,q-1)=1$. Then, 
\begin{equation*}
g(x)=x^{(q+1)l+(p-1)q+1}+x^{(q+1)l+pq}-x^{(q+1)l+q+(p-1)}
\end{equation*}
is a permutation polynomial of $\mathbb{F}_{q^2}$ if and only if $k$ is even. 
\end{theorem} 
{\em Proof:} We can rewrite $g(x)$ as $$x^{(q+1)l+p}\left(x^{(p-1)(q-1)}+x^{p(q-1)}-x^{q-1}\right).$$
Note that $\gcd((q+1)l+p,q-1)=\gcd(2l+p,q-1)=1$. Then, by Lemma \ref{zieve lemma}, $g(x)$ is a permutation polynomial of $\mathbb{F}_{q^2}$ if and only if $x^{(q+1)l+p}\left(x^{(p-1)}+x^{p}-x\right)^{q-1}$ permutes $U$. The latter is equivalent to that $x^{p}\left(x^{(p-1)}+x^{p}-x\right)^{q-1}$ permutes $U$ since $x^{(q+1)l}=1$ when $x\in U$. The desired result follows now from Corollary \ref{coro1}.
  \hfill$\square$
 
 \begin{remark}As we have shown in the proofs of Corollary \ref{coro1} and Theorem \ref{permutation property of g(x)},  $f(x)$ or $g(x)$ is a permutation polynomial of $\mathbb{F}_{q^2}$ if and only if  $x^{p}\left(x^{(p-1)}+x^{p}-x\right)^{q-1}$ permutes $U$. This shows that the permutation property of $f(x)$ and that of $g(x)$ depend on a same condition. However, it seems difficult to verify  directly whether this condition holds.
Thus, in this paper we use a different approach to investigate the permutation property of $f(x)$, and then obtain the permutation property of $g(x)$. \end{remark}
 
 \begin{remark}
 Let $f(x)$ and $g(x)$ be defined in (\ref{niho4-per}) and (\ref{general-niho-per}), respectively. The polynomial $g(x)$ contains $f(x)$ as a special case, and by taking  $l=0$, $g(x)$ is transformed into exactly  $f(x)$.  Note that when $p=2$, $f(x)=x^{pq}$, $g(x)=x^{(q+1)l+pq}$ and they are always permutation polynomials of $\mathbb{F}_{2^{2k}}$ for any positive integer $k$. For $p>5$, Theorems \ref{main result} and \ref{permutation property of g(x)} may not hold. By Magma, we have obtained some numerical results in Table \ref{tab1}.

\begin{table}\label{tab1}\caption{Is $f(x)$ or $g(x)$ a permutation over $\mathbb{F}_{p^{2k}}$?}
 \begin{center}
 \begin{tabular}{|c|c|c|}
 \hline 
  $p$& $k$ &  Permutation over $\mathbb{F}_{p^{2k}}$\\ 
  \hline 
    $7$ & $1$ &  No \\ 
 \hline 
  $7$ & $2$ &  Yes \\ 
  \hline
   $7$ & $3$ &  No \\ 
   \hline
    $7$ & $4$ &  No \\ 
 \hline 
  $7$ & $5$ &  No \\ 
    \hline
     $7$ & $6$ &  No \\ 
   \hline 
    $11$ & $1$ &  No \\ 
      \hline 
      $11$ & $2$ &  Yes \\ 
         \hline 
       
       $11$ & $3$ &  No \\   
      \hline 
      $11$ & $4$ &  No \\   
     
  \hline 
 $13$ & $1$ &  No \\ 
   \hline 
   $13$ & $2$ &  Yes \\ 
      \hline 
    
    $13$ & $3$ &  No \\ 
      \hline

 \end{tabular} 
 \end{center}
 \end{table}

 \end{remark}
 
 \section{A comparison with known related permutation trinomials}\label{sec4}
 
In this section, we will compare the permutation polynomials $f(x)$ proposed in Theorem \ref{main result} with previously  known ones of the form (\ref{niho0-per}). It is straightforward that  the composition of two permutation polynomials of the same finite field is also a permutation polynomial. 
We recall the definition of multiplicative equivalence from \cite{Hou2015survey,Likangquan, wudanyao2017}.

\begin{definition}
Let $q$ be a prime power, and  $H(x)$ and $h(x)$ be two permutation polynomials of $\mathbb{F}_q$.  $H(x)$ and $h(x)$ are called multiplicative equivalent if there exists an integer $1\leq d\leq q-2 $ such that $\gcd(d,q-1)=1$ and $H(x)=ah(x^d)$ for some $a\in{\mathbb F}_{q}^*$.

\end{definition}

%In this section, we will show that the permutation trinomials $f(x)$ presented in Theorem \ref{main result}
%are not multiplicative equivalent to the permutation trinomials of similar form.  We mainly compare our permutation trinomials $f(x)$ with the ones of the form (\ref{niho0-per}) since they are closely related to each other.  Precisely, 

Next we will determine whether or not the permutation trinomials $f(x)$ given in Theorem \ref{main result} are  multiplicative equivalent to the previously known ones of the form (\ref{niho0-per}) in \cite{Likangquan2016arxiv,Linian2016conjecture,Wu-Li2017,mage2017,X.Hou-2014method,X.Hou-2014arxiv,
Ding-siam,Kyureghyan-Zieve2016}. We make some preparations as follows.

\begin{proposition}\label{multi-equi}
Let $f(x)$ be defined in (\ref{niho4-per}), and $q=p^k$ with $k$ being even and $p \in \{3,5\}$. We have the following results:

\noindent{(a)} If $p=3$, then $f(x)$ is multiplicative equivalent to the following permutation trinomials of $\mathbb{F}_{q^2}$:

{(i)} $f_1(x)=x+x^{(2\cdot 3^{k-1}+1)(q-1)+1}-x^{(3^{k-1}+1)(q-1)+1}$;

{(ii)} $f_2(x)=x+x^{-q+2}-x^{q}$;

{(iii)} $f_3(x)=x-x^{q}-x^{2q-1}$;

\noindent{(b)} If $p=5$, then $f(x)$ is multiplicative equivalent to the following permutation trinomials of $\mathbb{F}_{q^2}$:

{(i)} $f_1(x)=x+x^{(4\cdot 5^{k-1}+1)(q-1)+1}-x^{(5^{k-1}+1)(q-1)+1}$;

{(ii)} $f_2(x)=x+x^{\frac{2q+1}{3}(q-1)+1}-x^{q}$;

{(iii)} $f_3(x)=x-x^{q}-x^{\frac{q+5}{3}(q-1)+1}$.
\end{proposition}
{\em Proof:} We only prove the case $p=5$. For the case $p=3$, the result can be proved in the same way.  When $p=5$, note that  $f_1(x)=f(x^{5^{k-1}})$. Thus, $f(x)$ is multiplicative equivalent to $f_1(x)$. Let $d_1=(4\cdot 5^{k-1}+1)(q-1)+1$ and $d_2=(5^{k-1}+1)(q-1)+1$. When $k$ is even, $\gcd(d_1,q^2-1)=\gcd(d_2,q^2-1)=1$. Then, by extended Euclidean algorithm, we have $d_1^{-1}=\frac{2q+1}{3}(q-1)+1$ and $d_2^{-1}=\frac{q+5}{3}(q-1)+1$, where $d_i^{-1}$ denotes is the inverse of $d_i$ modulo $q^2-1$, $i=1,2$.
Note that  $f_2(x)=f_1(x^{d_1^{-1}})$ and $f_3(x)=-f_1(x^{d_2^{-1}})$.
Therefore, $f(x)$ is multiplicative equivalent to $f_2(x)$ and $f_3(x)$. It is easily seen that $f_1(x)$, $f_2(x)$ and $f_3(x)$ are pairwise multiplicative equivalent to each other. 
 \hfill$\square$

The following claims are needed. 

{\em \textbf{Claim 1:}} Recall that the inverse $d^{-1}$ of a (normalized) Niho exponent $d$ over $\mathbb{F}_{p^{2k}}$, if it exists, is again a (normalized) Niho exponent, and the product of two (normalized) Niho exponents is also a (normalized) Niho exponents \cite{Dobbertin06}.

Let $F(x)$ be a permutation polynomial of $\mathbb{F}_{p^{2k}}$ of the form (\ref{niho0-per}), $d_1=s(p^k-1)+1$ and $d_2=t(p^k-1)+1$. If one of $d_1$  and $d_2$ is invertible, say $d_1$, then $\lambda_1 F(x^{d_1^{-1}})$ is also a permutation polynomial of the form (\ref{niho0-per}) which is multiplicative equivalent to $F(x)$. This analysis together with Claim 1 gives the following claim. 

{\em \textbf{Claim 2}:} Let $F(x)$ be a permutation polynomial of $\mathbb{F}_{p^{2k}}$ of the form (\ref{niho0-per}). Then, all the permutation trinomials of the form (\ref{niho0-per}) that are multiplicative equivalent to $F(x)$ are given by $\lambda_i F(x^{d_i^{-1}})$ provided $\gcd(d_i,p^{2k}-1)=1$, $i=1,2$.

Claims 1 and 2 together with Proposition \ref{multi-equi} give the following claim. 

{\em \textbf{Claim 3:}}  Let $f(x)$ be a permutation polynomial in Theorem \ref{main result}. If a permutation trinomial of the form (\ref{niho0-per}) is multiplicative equivalent to $f(x)$, then it must be one of $f_1(x), f_2(x)$ and $f_3(x)$ in Proposition \ref{multi-equi}.

 In the sequel, a permutation polynomial $F(x)$ of $\mathbb{F}_{p^{2k}}$ of the form (\ref{niho0-per}) will be denoted by the tuple $\left(\lambda_1, s,\lambda_2, t\right)$. According to Lemma \ref{zieve lemma}, $F(x)$ is a permutation polynomial of $\mathbb{F}_{p^{2k}}$ 
 if and only if the associated polynomial \begin{equation}\label{fractional poly}
  x\left(1+\lambda_1x^s+\lambda_2 x^t\right)^{p^k-1}\end{equation} permutes the unit circle $U$. When $F(x)$ is a permutation polynomial of $\mathbb{F}_{p^{2k}}$, (\ref{fractional poly}) can be further written as a fractional polynomial 
 \begin{equation*}
x\frac{1+x^{-s}+x^{-t}}{1+x^s+x^t}
 \end{equation*}
 since $x\in U$, which is called the fractional polynomial of $F(x)$. For comparison purposes, we collect all known permutation trinomials of $\mathbb{F}_{3^{2k}}$ and $\mathbb{F}_{5^{2k}}$ of the form (\ref{niho0-per}). We list them in Tables \ref{table 2} 
and \ref{table 1}, respectively.  To the best of our knowledge,  Tables \ref{table 2} 
and \ref{table 1} contain such permutation trinomials completely. 

Note that the last one in Table \ref{table 2} is exactly $f_3(x)$ in Proposition \ref{multi-equi} (a)(iii). In \cite{X.Hou-2014arxiv}, Hou determined all permutation trinomials of $\mathbb{F}_{q^2}$ of the form $ax+bx^q+x^{2q-1}\in \mathbb{F}_{q^2}[x]$. When $p=3$, according to Theorem A (iv) of \cite{X.Hou-2014arxiv}, $-x+x^q+x^{2q-1}$ is a permutation polynomial of $\mathbb{F}_{3^{2k}}$ if and only if $-1$ is a square of $\mathbb{F}_{3^{k}}^*$, which is equivalent to that $k$ is even. The permutation polynomial $x-x^q-x^{2q-1}$ equals $-\left(-x+x^q+x^{2q-1}\right)$, and thus its permutation property can be derived from Theorem A (iv) of \cite{X.Hou-2014arxiv}. 

\begin{table}[!t]\small
\renewcommand{\arraystretch}{1.3}
\caption{Known permutation trinomials of $\mathbb{F}_{3^{2k}}$ of the form  (\ref{niho0-per}) ($q=3^k$)} \label{table 2} \centering
\begin{tabular}{|c|c|c|c|c|}
\hline $\left(\lambda_1, s,\lambda_2, t\right)$ & Fractional polynomial& $k$&  Ref.\\

\hline $\left(-1, 2,1, -2\right)$ & $\frac{x^5+x^3-x}{-x^4+x^2+1}$& $k\not\equiv0\,\,(\mbox{mod}\,\,4)$&  \cite[Theorem 3.2]{Likangquan2016arxiv}\\

\hline $\left(1, 3,-1, -1\right)$ & $\dfrac{-x^4+x^3+1}{x^5+x^2-x}$& odd $k$ & \cite[Theorem 3.4]{Likangquan2016arxiv}\\

\hline $\left(-1, 4,1,-2\right)$ & $\dfrac{x^6+x^4-1}{-x^7+x^3+x}$& all $k$
 & \cite[Conjecture 5.1 (2)]{Likangquan2016arxiv}, \cite{Linian2016conjecture}\\

\hline $\left(-1, -2, 1, 2\right)$ & $\dfrac{-x^5+x^3+x}{x^4+x^2-1}$&  $k\not\equiv2\,\,(\mbox{mod}\,\,4)$& \cite[Conjecture 5.1 (3)]{Likangquan2016arxiv}, \cite{Linian2016conjecture}\\

\hline $\left(-1, \frac{q+3}{4}, -1, \frac{3q+5}{4}\right)$ & $x\dfrac{x^{\frac{3q+5}{4}}-x^{\frac{q+1}{2}}-1}{x^{\frac{3q+5}{4}}-x-x^{\frac{q+3}{2}}}$&  even $k$& \cite[ Theorem 1.1 (d)]{Kyureghyan-Zieve2016}\\

\hline $\left(-1, \sqrt{q}, -1, 1-\sqrt{q}\right)$ & $\dfrac{x^{\sqrt{q}}-x^{2\sqrt{q}-1}-1}{x^{\sqrt{q}-1}-x^{2\sqrt{q}-1}-1}$&  even $k$& \cite[ Theorem 1.1 (e)]{Kyureghyan-Zieve2016}\\

\hline $\left(-1, \sqrt{q}+1, -1, -\sqrt{q}\right)$ & $\dfrac{x^{\sqrt{q}+1}-x^{2\sqrt{q}+1}-1}{x^{\sqrt{q}}-x^{2\sqrt{q}+1}-1}$&  even $k$& \cite[ Theorem 1.1 (f)]{Kyureghyan-Zieve2016}\\

\hline
 $\left(-1, 1, -1, 2\right)$ & $\dfrac{x+1-x^2}{x^3+x^2-x}$&  even $k$& \cite[ Theorem A (iv)]{X.Hou-2014arxiv}\\
\hline
\end{tabular}
\end{table}

\begin{table}[!t]\small
\renewcommand{\arraystretch}{1.3}
\caption{Known permutation trinomials of $\mathbb{F}_{5^{2k}}$ of the form  (\ref{niho0-per}) ($q=5^k$)} \label{table 1} \centering
\begin{tabular}{|c|c|c|c|c|}
\hline $\left(\lambda_1, s,\lambda_2, t\right)$ & Fractional polynomial& $k$&  Ref.\\

\hline $\left(1, \frac{q+3}{4},-1, \frac{q+3}{2}\right)$ & $-x^{\frac{q+1}{2}}\left(\frac{x^s-2}{x^s+2}\right)^2(s=\frac{q+3}{4})$& all $k$&  \cite[Theorem 1]{Wu-Li2017}\\

\hline $\left(1, \frac{q-1}{2},-1, \frac{q+3}{2}\right)$ & $\dfrac{-x\left(x^{\frac{q+3}{2}}-x^{\frac{q-1}{2}}+1\right)}{x^{\frac{q+3}{2}}-x^{\frac{q-1}{2}}-1}$& odd $k$ & \cite[Theorem 2]{Wu-Li2017}\\

\hline $\left(1, -1,-1, \frac{q+3}{2}\right)$ & $\dfrac{x^2\left(x^{\frac{q-1}{2}}-x-1\right)}{x^{\frac{q+5}{2}}-x-1}$& odd $k$
 & \cite[Theorem 3]{Wu-Li2017}\\

\hline $\left(-1, \frac{q+3}{2}, 1, \frac{q+5}{2}\right)$ & $\dfrac{-x\left(x^{\frac{q-1}{2}}-x^{\frac{q-3}{2}}-1\right)}{x^{\frac{q+5}{2}}-x^{\frac{q+3}{2}}+1}$& odd $k$& \cite[Theorem 4]{Wu-Li2017}\\

\hline $\left(-1, 2, 1, \frac{q+3}{2}\right)$ & $\dfrac{x^{\frac{q+3}{2}}+x^2-1}{x\left(x^{\frac{q+3}{2}}-x^2+1\right)}$& even $k$& \cite[Theorem 5]{Wu-Li2017}\\

\hline $\left(1, 1, -1, \frac{q-1}{2}\right)$ & $\dfrac{x^{\frac{q+5}{2}}-x-1}{x^{\frac{q-1}{2}}-x-1}$& even $k$& \cite[Theorem 6]{Wu-Li2017}\\

\hline $\left(-1, 1, 1, \frac{q+5}{2}\right)$ & $\dfrac{x^{\frac{q-1}{2}}+x-1}{x^{\frac{q+5}{2}}-x+1}$& even $k$& \cite[Theorem 7 (i)]{Wu-Li2017}\\

\hline $\left(1, \frac{q+3}{2}, 1, \frac{q+5}{2}\right)$ & $\dfrac{x^{\frac{q+1}{2}}+x^{\frac{q-1}{2}}+x}{x^{\frac{q+5}{2}}+x^{\frac{q+3}{2}}+1}$& even $k$& \cite[Theorem 7 (ii)]{Wu-Li2017}\\

\hline $\left(1, \frac{q+3}{2}, -1, -1\right)$ & $\dfrac{x^2\left(x^{\frac{q-1}{2}}-x+1\right)}{x^{\frac{q+5}{2}}+x-1}$& even $k$ & \cite[Theorem 7 (iii)]{Wu-Li2017}\\

\hline $\left(1, 2, -1, -2\right)$ & $-x\left(\dfrac{x^2+2}{x^2-2}\right)^2$& odd $k$ & \cite[Proposition 1]{Wu-Li2017}\\ 
&&&\cite[Theorem 4.1]{mage2017}\\

\hline $\left(-1, \frac{q+5}{3}, -1, \frac{2(q+2)}{3}\right)$ & $-x\left(\dfrac{x^2-2}{x^2+2}\right)^2$& even $k$& \cite[Proposition 2]{Wu-Li2017}\\ 
&&&\cite[Theorem 3.1]{mage2017}\\

\hline $\left(1, \frac{q+2}{3}, 1, \frac{2q+4}{3}\right)$ & $\dfrac{x^{\frac{2q+4}{3}}+x^{\frac{q+2}{3}}+1}{x^{\frac{2q+1}{3}}+x^{\frac{q+2}{3}}+1}$& even $k$& \cite[ Theorem 1.1 (c)]{Kyureghyan-Zieve2016}\\

\hline $\left(-1, \sqrt{q}, -1, 1-\sqrt{q}\right)$ & $\dfrac{x^{\sqrt{q}}-x^{2\sqrt{q}-1}-1}{x^{\sqrt{q}-1}-x^{2\sqrt{q}-1}-1}$&  even $k$& \cite[ Theorem 1.1 (e)]{Kyureghyan-Zieve2016}\\

\hline $\left(-1, \sqrt{q}+1, -1, -\sqrt{q}\right)$ & $\dfrac{x^{\sqrt{q}+1}-x^{2\sqrt{q}+1}-1}{x^{\sqrt{q}}-x^{2\sqrt{q}+1}-1}$&  even $k$& \cite[ Theorem 1.1 (f)]{Kyureghyan-Zieve2016}\\

\hline $\left(1, t, 1, -t\right)$ & $x$&  even $k$& \cite[ Theorem 3.4 (i) ]{Ding-siam}\\

\hline $\left(1, t, 1, -t\right)$ & $x$& $\begin{array}{cc}\mbox{odd}\,\, k\,\,\mbox{and}\\ {\rm{exp}}_3(t)\geq {\rm {exp}}_3(q+1)^*\end{array}$& \cite[ Theorem 3.4 (iii) ]{Ding-siam}\\

\hline $\left(\pm 1, 1, 1, 2\right)$ & $\dfrac{1}{x}$& even $k$& \cite[ Theorem A (ii) ]{X.Hou-2014method}\\

\hline
\end{tabular}
$*$where ${\rm {exp}}_3(i)$ denotes the exponent of $3$ in the canonical factorization of $i$.~~~~~~~~~~~~~~~~~~~~~~~~~~~~~~~~~~~~~~~~~~
\end{table}

According to Tables \ref{table 2}-\ref{table 1}, Proposition \ref{multi-equi} and Claim 3, we conclude the following result.

\begin{proposition}\label{not equivalent}  Let $f(x)$ be a permutation polynomial proposed in Theorem \ref{main result}. When $p=3$, $f(x)$ is multiplicative equivalent to the permutation polynomial $x-x^q-x^{2q-1}$ (or $-x+x^q+x^{2q-1}$ ) which is contained in Theorem A (iv) of \cite{X.Hou-2014arxiv}. When $p=5$, $f(x)$ is not multiplicative equivalent to any permutation trinomial listed in Table \ref{table 1}. 
 \end{proposition}

 The above proposition shows that  $f(x)$ proposed in Theorem \ref{main result} is indeed new when $p=5$. When $p=3$, the permutation polynomial $f(x)$ proposed in Theorem \ref{main result} is multiplicative equivalent to a known one contained in \cite{X.Hou-2014arxiv}. Nevertheless, the method for proving the permutation property in this paper are different from that in \cite{X.Hou-2014arxiv}.

\section{Conclusion}\label{sec5}
In this paper, we construct a class of permutation trinomials of $\mathbb{F}_{q^2}$ with $q=3^k$ and $5^k$. Precisely, for each $p\in \{3,5\}$, we prove that $f(x)=x^{(p-1)q+1}+x^{pq}-x^{q+(p-1)}$ is a permutation trinomial of $\mathbb{F}_{q^2}$  if and only if $k$ is even. This conclusion is also true for more general polynomials $g(x)=x^{(q+1)l+(p-1)q+1}+x^{(q+1)l+pq}-x^{(q+1)l+q+(p-1)}$ with $l$ being a nonnegative integer satisfying $\gcd(2l+p,q-1)=1$. Moreover,  when $p=5$, we prove that $f(x)$ presented here is  not multiplicative equivalent to any known permutation trinomial of the form (\ref{niho0-per}).  Numerical experiments show that Theorems \ref{main result} and \ref{permutation property of g(x)} may not hold when $p>5$. It would be nice if  our construction can be generalized to arbitrary finite field. Namely, the readers are invited to determine the permutation polynomials of $\mathbb{F}_{q^2}$ of the form  
$$x^{(p-1)q+1}+\lambda_1x^{pq}+\lambda_2x^{q+(p-1)},$$
where $p$ is a prime, $q=p^k$ for some positive integer $k$ and 
$\lambda_1,\lambda_2\in \mathbb{F}_{q}$.

\section*{Acknowledgment}
T. Bai and Y. Xia were supported in part by National Natural Science Foundation of China under Grant 61771021 and Grant 11301552, and in part by Natural Science Foundation of Hubei Province under Grant 2017CFB425. T. Bai was also supported by Graduate Innovation Fund of South-Central University for Nationalities.

\end{document}